# Individual Particle Localization per Relativistic de Broglie-Bohm

David L. Bartley
Consultant, Physics and Statistics, 3904 Pocahontas Avenue, Cincinnati, OH 45227, USA

**Abstract**

The significance of the de Broglie-Bohm hidden-particle position in the relativistic regime is addressed, seeking connection to the (orthodox) single-particle Newton-Wigner position. The effect of non-positive excursions of the ensemble density for extreme cases of positive-energy waves is easily computed using an integral of the equations of motion developed here for free spin-0 particles in 1+1 dimensions and is interpreted in terms of virtual-like pair creation and annihilation beneath the Compton wavelength. A Bohm-theoretic description of the acausal explosion of a specific Newton-Wigner-localized state is presented in detail. The presence of virtual pairs found is interpreted as the Bohm picture of the spatial extension beyond single point particles proposed in the 1960s as to why space-like hyperplane dependence of the Newton-Wigner wavefunctions may be needed to achieve Lorentz covariance. For spin-1/2 particles the convective current is speculatively utilized for achieving parity with the spin-0 theory. The spin-0 improper quantum potential is generalized to an improper stress tensor for spin-1/2 particles.

**Keywords**

Dirac; de Broglie; Bohm; relativistic, Newton-Wigner; acausal; spin tensor; stress; pair creation; particle

# Introduction

A puzzling aspect of "relativizing" the de Broglie/Bohm (dBB) deterministic model[1-5] of individual particles is whether the sub-quantum particle trajectories, the model's hidden variables, retain *any* significance. Within the non-relativistic theory, averages over particle ensembles agree with predictions of orthodox quantum mechanics. However, the covariant sub-quantum particle positions that emerge from the Gordon-Klein-Schroedinger (GKS) theory relate directly to the wavefunction's spatial configuration coordinates which are not directly observable as they are eigenvalues of no Hermitean operator.

Following Newton and Wigner[6] (NW), we are led to consider the possible connection of sub-quantum trajectories to the eigenvalues (the NW positions) of the Hermitean part of the position operators, even though in their original form not manifestly covariant. The NW position has been found[7] easily made covariant by relaxing what may be a preconception that the particles always appear as strictly single point particles. Such a spatial extension may occur only in extreme conditions that have not yet been subject to experiment.

The GKS theory of positive-energy particles has a conserved density that, unlike the non-relativistic theory may, in some cases be negative, like a charge density, rather than probability density. This suggests that the theory may describe how extreme circumstances may result in the presence of an even number of virtual particle/anti-particle pairs, the anti-particles defined in terms of the Feynman/ Stückelberg speculation[8-10] as particles moving backwards in time. The pairs are only similar to the more complete multi-particle theory covered by quantum field theory where interactions effect particle creation and annihilation. In fact, creation and annihilation in the free-particle theory are not Lorentz-invariant events, but are the relativistic manifestation of the arbitrarily large fluctuations in the particle velocity possible in the non-relativistic dBB theory.

The separation between such pairs could then explain the spatial extension of what in the non-relativistic theory are strictly single point particles. Any pairs created within the interaction-free theory are comprised of *virtual* particles that are "hyper-entangled" in the sense that they recombine within only short times—of the order of the (Compton wavelength) / (speed of light). The virtual anti-particles are found only at the sub-quantum level, never at the level of orthodox single-particle quantum mechanics. This paper explores the possibility that the NW position at given time may be understood as a function of the hidden particle and/or pair trajectories in extreme circumstances.

The plan of the paper is to first briefly describe the spin-0 dBB theory which directly leads to an integral of the equations of motion for 1+1 dimensions, greatly simplifying the calculation of explicit particle trajectories. The NW theory (*orthodox* in the sense of initially ignoring dBB) is introduced via wavefunctions of the NW position in terms of the covariant solutions to the force-free GKS equation. Controversies over the NW position itself are briefly described, namely stemming from the a-covariant nature and the apparent superluminal expansion from an initial spatial confinement discovered several years ago.[7] Resolution of controversy in acquiring covariance by acknowledging the possibility of departure from strictly individual point particles is described in terms of allowing the NW wavefunction to depend explicitly on the space-like hyperspace of particle observation at given time. Near non-relativistic reproduction of the NW position in terms of Bohmian sub-quantum positions (0-th order) perturbed (at 1-st order) by dependence on the pilot field local to particle trajectory is then developed. A dBB-theoretic description of the acausal explosion of a specific Newton-Wigner-localized state is then covered in detail. The presence of virtual pairs found is interpreted as the dBB picture of spatial extension beyond single point particles.

**The Scalar Field and de Broglie/Bohm, Introduction**

The basics behind a relativistic de Broglie/Bohm theory are presented here for individual Gordon-Klein-Schrödinger (GKS) particles. Many researchers have published in this area; see Ref. 11 and references within. For more recent work, also see Refs. 12-15 and referenced papers.

Let $\psi$ be a solution to the force-free GKS equation:

$$-\hbar^2 \,\Box\, \psi = -(m_0 c)^2 \psi, \tag{1}$$

where the symbol $\Box$ represents the D'Alembertian operator. Representation of arbitrary complex scalar $\psi$ in terms of amplitude $A$ and phase $S$ as

$$\psi = A e^{\frac{i}{\hbar} S} \tag{2}$$

leads to a conserved density $\rho$ via current $J_\mu$,

$$\begin{aligned} J_\mu &= \frac{\hbar}{2i} \left[ \psi^* \frac{\partial \psi}{\partial x_\mu} - \frac{\partial \psi^*}{\partial x_\mu} \psi \right] \\ &= A^2 \partial S / \partial x_\mu, \end{aligned} \tag{3}$$

The density $\rho$, not necessarily positive, is interpreted here as proportional to the difference between ensemble densities of particles and anti-particles. Furthermore a relativistic version of the Hamilton-Jacobi equation follows:

$$\partial_\nu S \,\partial^\nu S - \hbar^2 A^{-1} \,\Box\, A = -(m_0 c)^2. \tag{4}$$

A classical relativistic mechanics is then obtained by identifying the momentum $p_\mu$ as:

$$p_\mu = \partial S / \partial x_\mu$$
$$= \mu_0(x) u_\mu, \qquad (5)$$

where $u_\mu$ is the proper velocity, and $\mu_0(x)$ is a function which may be regarded as an effective invariant mass (avoiding the term "rest mass") of particle in the presence of "pilot wave". Allowance for an effective mass is suggested by the simplest relativistic mechanics described by the variational form:

$$\delta \int d\tau \, \mu_0 c^2 = 0, \qquad (6)$$

where $d\tau$ is a proper time interval.

As $u_\mu u^\mu = -c^2$, Equations (4-5) imply that defining a scalar $\Phi$ by:

$$\Phi \equiv -\tfrac{\hbar^2}{2m_0} A^{-1} \, A, \qquad (7)$$

the effective invariant mass $\mu_0(x)$ is given by:

$$(\mu_0 c)^2 = (m_0 c)^2 + 2 m_0 \Phi. \qquad (8)$$

Equation (8) may require expansion for the spin-1/2 particles as indicated in Appendix B. As seen below, some limited particle trajectories may have $|v| > c$ for brief periods, where the invariant mass is imaginary; however, the energy or relativistic mass ($\mu_0(x)/\sqrt{1-(v/c)^2}$ from $p_4$) remains finite and real even at the passage of $\sqrt{1-(v/c)^2}$ through zero.

Note that adopting a sign convention for the real or imaginary component of $\sqrt{1-(v/c)^2}$ as negative in the antiparticle region (where $\rho < 0$) is useful in leading to real and imaginary parts of the proper time increasing along trajectories, and also the relativistic mass becomes negative at the boundary $\rho = 0$ with continuous derivatives. The momentum $p$ and $v$ are then opposite at $\rho < 0$.

The scalar $\Phi$ can be identified as the quantum potential, since differentiating the Hamilton-Jacobi Equation (4), noting that $\partial p_\upsilon / \partial x_\mu = \partial p_\mu / \partial x_\upsilon$, gives:

$$\mu_0 u^\upsilon \frac{\partial(\mu_0 u_\mu)}{\partial x_\upsilon} = -\partial(m_0 \Phi)/\partial x_\mu$$
$$= \mu_0 \frac{d(p_\mu)}{d\tau} . \qquad (9)$$

Equation (9) provides a description of a fluid-like development of an ensemble of particles with proper velocity $u$ at space-time $x$.

Finally, Equation (5) implies that the particle velocity $v_j$ is given by:

$$v_j = J_j / \rho$$
$$= -c^2 \frac{\partial S/\partial x_j}{\partial S/\partial t}$$
$$= -c^2 \frac{\left[\psi^* \dfrac{\partial \psi}{\partial x_j} - \dfrac{\partial \psi^*}{\partial x_j}\psi\right]}{\left[\psi^* \dfrac{\partial \psi}{\partial t} - \dfrac{\partial \psi^*}{\partial t}\psi\right]}. \tag{10}$$

**Integral of the Equations of Motion in 1+1 Dimensions**

In 1+1 dimensions, the computation of sub-quantum particle trajectories given by Equation (10) is vastly simplified because of the existence of an integral of the equations of motion. If the wave function $\psi$ is a superposition of plane waves in only one spatial dimension, Equation (10) can be integrated explicitly so that particle trajectories can be easily determined. Integration is simple in this case because the frequencies $\omega$ for any two wave-number values $k$ and $k'$ satisfy:

$$\omega_{k'}^2 - \omega_k^2 = c^2(k'^2 - k^2). \tag{11}$$

Then an integral $I$ of the equations of motion, with only a single spatial dimension, is given by:

$$I = \int \frac{dk'}{\omega_{k'}} \int \frac{dk}{\omega_k} \frac{\omega_k + \omega_{k'}}{k' - k - i\eta} \phi_{k'}^* \phi_k \, e^{i(k-k')z - i(\omega_k - \omega_{k'})t}, \tag{12}$$

where

$$\psi[z,t] = \int \frac{dk}{\omega_k} \phi_k \, e^{i(kz - \omega_k t)}. \tag{13}$$

That $I$ is constant follows directly from Equation (10) by computing $dI/dt$, regarding $z$ as a function of time $t$, noting that the term $i\eta$ in Equation (12) simply adds (via Sokhotsky's formula) a constant (independent of $z$ and $t$) into $I$.

The integral *I* is used later in this paper for computing trajectories. A version of *I* for discrete plane waves is also useful, illustrating the general nature of virtual pair creation and annihilation in the case of extreme states. See Appendix A for details.

**Measurement: Newton-Wigner Position**

Contact with the statistical predictions of orthodox quantum theory is considered here. The Newton-Wigner (NW) position[6] is shown in the lowest nontrivial relativistic limit to be representable in terms of the Bohmian trajectory together with aspects of the field local to the trajectory. The Bohmian position is correspondingly evidently less open to measurement within quantum mechanical limitations than in the nonrelativistic theory.

**introduction and notation**

We are concerned here with positive solutions of the free spin-0 Gordon-Klein-Schrödinger (GKS) which we consider consistent mathematically, even though not representative of multi-particle systems and whether or not closely relevant to nature.

With an arbitrary scalar (spin-0) positive-energy wavefunction $\psi[x]$ denoted as:

$$\psi[x] = \int \frac{d^3k}{\omega_k (2\pi)^{3/2}} \phi_k e^{i(k \cdot x)}, \tag{14}$$

the inner product is given in momentum space by:

$$(\psi_2, \psi_1) = \int \frac{d^3k}{\omega_k} \phi_{2k}^* \varphi_{1k}. \tag{15}$$

The Newton-Wigner position, denoted here as $\vec{x}$, is an eigenvalue of the operator $\hat{\vec{x}}$, defined by:

$$\hat{\vec{x}} = i\partial/\partial k_j - \tfrac{1}{2} i k_j / \omega_k^2, \tag{16}$$

which is the Hermitean part of $\vec{x} = i\partial/\partial k$, considering the above inner product. The frequency $\omega_k$ is

$$\omega_k = \sqrt{1 + \vec{k} \cdot \vec{k}} \quad \text{(at } \hbar, c, \text{ and } m_0 \text{ set equal to 1).} \tag{17}$$

The corresponding momentum-space eigenfunction with NW position eigenvalue $\vec{x}$ is:

$$\phi_{\underline{x}} = (2\pi)^{-3/2} \omega_k^{1/2} e^{-ik \cdot \underline{x}} \quad (k \equiv (\omega_k, \vec{k})). \tag{18}$$

Therefore,

$$(\varphi_{\underline{x}}, \varphi_{\underline{x}'}) = \delta^{(3)}[\vec{\underline{x}} - \vec{\underline{x}}'] \quad at \quad \underline{t} = \underline{t}', \tag{19}$$

though neither $\psi_{\underline{x}}[x]$ nor $\partial \psi_{\underline{x}}[x]/\partial t$ are delta functions in coordinate space.

**Newton-Wigner position density**

The wavefunction $\underline{\psi}[\underline{x}]$ for measuring $\vec{\underline{x}}$, given the general $\psi[x]$ above is then:

$$\begin{aligned} \underline{\psi}[\underline{x}] &= (\psi_{\underline{x}}, \psi) \\ &= \int \frac{d^3k}{\omega_k (2\pi)^{3/2}} \omega_k^{1/2} \varphi_k \, e^{i(k\cdot \underline{x})}. \end{aligned} \tag{20}$$

$|\underline{\psi}[\underline{x}]|^2$ is the (positive definite) probability density $\rho_{NW}[\vec{\underline{x}}, \underline{t}]$ for the NW position $\vec{\underline{x}}$.

Suppose the system, prior to measurement, is specified as $\underline{\psi}[\underline{x}]$, resulting in the Newton-Wigner density given by

$$\begin{aligned} \rho_{NW}[\vec{\underline{x}}, \underline{t}] &= \left|\underline{\psi}[\underline{x}]\right|^2 \\ &= \underline{\omega}^{1/2} \underline{\psi}^*[\underline{x}] \cdot \underline{\omega}^{1/2} \underline{\psi}[\underline{x}], \end{aligned} \tag{21}$$

where the operator $\underline{\omega}$ is defined as

$$\underline{\omega} \equiv c[1-\Delta]^{1/2} \tag{22}$$

for expansion in the Laplacian operator $\Delta$.

$\rho_{NW}[\vec{\underline{x}}, \underline{t}]$ is to be compared to the usual time-component $\rho[\vec{x}, t]$ of the 4-current defined here by:

$$\begin{aligned} \rho[x] &= \tfrac{i}{2}[\psi^*[x] \cdot \partial \psi[x]/\partial t - \partial \psi^*[x]/\partial t \cdot \psi[x]] \\ &= \tfrac{1}{2}[\psi^*[x] \cdot \underline{\omega}\psi[x] + \underline{\omega}\psi^*[x] \cdot \psi[x]] \end{aligned} \tag{23}$$

(for positive-energy wavefunctions). Note that $\rho[\vec{x},t]$, possibly negative, refers here to a "charge" density at the sub-quantum or Bohmian level, whereas $\rho_{NW}[\vec{\underline{x}}, \underline{t}]$ ($|\underline{\psi}[\underline{x}]|^2$) is positive-definite in describing a single-particle density at the orthodox level.

## controversy over the Newton-Wigner position

Equation (19) is consistent with an intuitive idea of position, and the Newton-Wigner position $\underline{\vec{x}}$ often closely approximates the non-relativistic value. However, Newton-Wigner position has non-intuitive aspects. Suppose that the wave function is localized to the extent that at $\underline{t} = 0$, the amplitude so obtained is non-zero only over a spatial region of $\underline{\vec{x}}$ with finite extent. Then (almost incredibly) at $\underline{t} > 0$, the density $\rho_{NW}$ is (generally "small" though) non-vanishing *for all* $\underline{\vec{x}}$ outside the light cone from any initial position.[7,16]

[With only one spatial dimension, this statement is most directly proved (and other integrals of interest computed) by distorting the amplitude's path of integration in the complex-k plane, simplified since the Fourier transform is a *proper integral* and therefore without poles and whose limits themselves limit behavior at infinity in the $k$ plane: at any time $t > 0$ in the case of $x$ in the space-like region, the result is a non-vanishing integral circling the branch cut of $\omega_k$ in one of the half $k$ planes.]

Similarly, an observer Lorentz-boosted observes a Lorentz-contracted density, but also sees tails with the space-filling aspect invisible initially to the other observer. The system acts as if a space-filling *something* exists initially that is invisible to only one of the observers as resulting from a space-time parallax. Note also that it is possible to show that the energy density is space-filling for both observers. Of course, how the Newton-Wigner position would be measured is not obvious, but some schemes would seem to require interaction involving strictly infinite energy to set up a localized eigenstate (18). This is a subject within the domain of quantum field theory beyond the scope of the present paper.

These seemingly bizarre observations were preceded by objections that the wavefunction $\underline{\psi}[\underline{x}]$ (20) is not manifestly covariant as noted, in fact, in the original paper[6] by Newton and Wigner. Because of the pesky $\omega_k^{1/2}$ in Equation (20), $\underline{\psi}[\underline{x}]$ does not transform as a scalar under Lorentz transformations. This fact and the prospect of superluminality led Wigner[17] to question the operator $\underline{\vec{x}}$ as a legitimate position observable.

As remarked by Wigner[17], a resolution to this problem has been proposed as an "elegant relativistic formulation by Fleming", who noted that classical (i.e., non-quantum) position variables characterizing spatially-*extended* observables require specification of the space-like hyperplane in which the system is observed.[7] Examples are relativistic versions of the center of energy or charge. In the quantum realm, a wavefunction $\underline{\psi}$ could then possibly depend not only on the space-time point $\underline{x}$, but also on hyperplane.

Once this leap in understanding is achieved, it is simple to achieve manifest covariance with the free spin-0 system considered here: With space-time parsed into parallel flat space-like hyperplanes, let $n^\mu$ represent a time-like unit 4-vector normal to the hyperplanes. Then $\omega_k^{1/2}$ in Equation (20) is generalized to $\sqrt{k_\mu n^\mu}$, which is a scalar that reduces to $\omega_k^{1/2}$ in the case that $n^\mu$ is aligned with the $x^0$ axis (i.e., if $n^0 = 1 \text{ and } \vec{n} = \vec{0}$).

Then $\phi_{\underline{x}}$ above (18) is generalized to:

$$\phi_{\underline{x}} = (2\pi)^{-3/2} \sqrt{k \bullet n} \ e^{-ik \cdot \underline{x}}. \tag{24}$$

This implies that the relation (19) only holds in the special condition that $n = (1, \vec{0})$. Given the generalized position eigenfunction, the corresponding general $\underline{\psi}[\underline{x}; n]$ (20) transforms as a scalar.

Violating (19) upon Lorentz transformation, it is simple to show that projection operators onto disjoint bounded space-like regions of the Newton-Wigner space $\underline{x}$, may commute at $n = (1, \vec{0})$, though their unitarily-transformed operators under many Poincaré transforms do not. Non-commutation was recently proved, considering Poincaré translations, for parallel hyperplanes under general assumptions by Malament, who infers that measurements in one space-time region could affect those in a space-like separated region.[18] On these grounds, Malament regards relativistic quantum mechanics of a localizable individual particle as impossible and that only quantum field theory makes sense.

Fleming points out[19], however, that *approximate* position measurements may be possible without violating causality even in the face of superluminality. Furthermore, there is no *a priori* reason that GKS generalization from non-relativistic single-particle quantum mechanics preserves isolated point particles. Peculiarities of particle positioning may simply represent unfamiliar or unwarranted preconceptions. The suggestion then is that, rather than isolated point particles, the particles of the GKS equation may in some situations at least be spatially extended in an unspecified way. Finally, the disjoint projection operators are associated with the fundamental field $\psi[x]$ connected with measurement but without compact support. Therefore, it may be premature to conclude that measurement within a compact region can actually be carried out (from the point of view of quantum field theory). For a more formal discussion, extension to arbitrary spin, examples of classical systems, formal development of state vectors in terms of hyperplane identified by time-like normal $n$ and invariant $\tau$ describing the time-like distance of a given hyperplane from origin, and other topics, see References (7 and 19). References (20 and 21) also answer others' objections to the hyperplane approach. In the following, we will be working solely in a "lab" coordinate system where $n = (1, \vec{0})$.

**Near-classical density**

In any case, the question arises as to how a measurement of $\vec{\underline{x}}$ at $\underline{t}$ relates to a specific trajectory of a particle at $\vec{x}$ as described by the Bohm theory. Clearly, no linear single-particle Hermitean position operator can correspond directly to the multiple eigenvalues values at specific $t$ as measurements of $\vec{\underline{x}}$ itself in the extreme cases of the presence of virtual pairs as in Appendix A. What is the meaning of the particle position in these trajectories?

Insight into position measurement can be obtained by comparing the distributions of $\vec{x}$ and $\underline{\vec{x}}$ in the lowest relativistic limit (replacing $\underline{x}$ by $x$ for comparing functional forms). Expanding $\underset{\sim}{\omega}$ and $\underset{\sim}{\omega}^{1/2}$ (22) to second order in $\Delta$, Equations 21 and 23 lead directly (with 0-th and 1-st orders canceling) to:

$$\rho[x] - \rho_{NW}[x] \approx \tfrac{1}{32}(\Delta^2\psi^*\psi - 2\Delta\psi^*\Delta\psi + \psi^*\Delta^2\psi). \tag{25}$$

Now the divergence of the current (3) is given by:

$$\operatorname{div} J = \frac{1}{2i}\left[\psi^*\Delta\psi - \Delta\psi^*\psi\right]. \tag{26}$$

With the Schrödinger equation (for positive-energy solutions) expanded as:

$$\begin{aligned} i\,\partial\psi/\partial t &= \underset{\sim}{\omega}\psi \\ &\sim (1 - \tfrac{1}{2}\Delta)\psi, \end{aligned} \tag{27}$$

the time derivative of $\operatorname{div} J$ is then approximated as:

$$\partial \operatorname{div} J/\partial t \sim \tfrac{1}{4}(\Delta^2\psi^*\psi - 2\Delta\psi^*\Delta\psi + \psi^*\Delta^2\psi). \tag{28}$$

Therefore,

$$(\rho - \rho_{NW})\big|_{\vec{x}} \approx +\tfrac{1}{8}\partial \operatorname{div} J/\partial t \tag{29}$$

$$= -\tfrac{1}{8}\partial^2\rho/\partial t^2. \tag{30}$$

Note that Equations (29 and 30, respectively) indicate that the approximate densities $\rho$ and $\rho_{NW}$ are equivalently normalized, and that the expected values of $\underline{\vec{x}}$ and $\vec{x}$ are equal, as with the exact densities.

**near-classical Newton-Wigner position**

Reproduction of the Newton-Wigner position $\underline{\vec{x}}$ in terms of a Bohmian trajectory can now be given. Let us approximate

$$\underline{x}_i = x_i + f_i, \tag{31}$$

for some function $f_i$ with $|f_i| \ll |x_i|$. Then the distribution $\rho[\vec{x}]$ can be found from:

$$\rho_{NW}[\underline{\vec{x}}]d^3\underline{x} = \rho_{NW}[\underline{\vec{x}}[\vec{x}]]\tfrac{\partial(\underline{x}_1,\underline{x}_2,\underline{x}_3)}{\partial(x_1,x_2,x_3)}d^3x = \rho[\vec{x}]d^3x, \tag{32}$$

where the Jacobian is linearized as:

$$\frac{\partial(x_1,x_2,x_3)}{\partial(\underline{x}_1,\underline{x}_2,\underline{x}_3)} \simeq 1 + \partial f_i / \partial x_i . \tag{33}$$

With $\rho_{NW}[\underline{\vec{x}}[\vec{x}]]$ linearized as:

$$\rho_{NW}[\underline{\vec{x}}[\vec{x}]] \approx \rho_{NW}[\vec{x}] + f_i \partial \rho[\vec{x}]/\partial x_i , \tag{34}$$

finally, using Equation (29), we have simply:

$$f_i \approx \tfrac{1}{8} \frac{1}{\rho} \frac{\partial(\rho v_i)}{\partial t} ; \tag{35}$$

$$\underline{x}_i \approx x_i + \tfrac{1}{8} \frac{1}{\rho} \frac{\partial(\rho v_i)}{\partial t} , \tag{36}$$

with corrections proportional to $c^{-2}$. Therefore, in this near-non-relativistic limit, the Newton-Wigner position $\underline{\vec{x}}$ is given in terms of Bohmian particle characteristics as well as the field local to the particle. In fact, as the operator $\underline{\omega}$ (22) expands to a series with spatial derivatives of all orders, it is likely that $\underline{x}_i$ itself expands beyond 1-st order in (36) also to all orders. The Newton-Wigner position at time t would then be local only in approximation and more exactly depends on the entire trajectory of the sub-quantum particle.

**Bohmian trajectories for acausally exploding initially localized states**

The particle trajectories corresponding to the acausally spreading Newton-Wigner position-probability density evolving from an initially localized NW wavefunction are very interesting and involve virtual particle/anti-particle annihilation as in Fig. A-1. In this ultra-relativistic case, the density $\rho$ may be negative (even for positive-energy wavefunctions); $\rho$ is considered a hidden-particle "charge" density, rather than probability density function. Trajectories are easily calculated using the integral of the equations of motion (Equation (12)) for one spatial dimension. As an example localized initial state, consider a specific wavefunction $\underline{\psi}[\underline{x}]$ (as in Equation (20)) for measuring Newton-Wigner position $\underline{x}$ at $\underline{t} = 0$:

$$\underline{\psi}[\underline{x}] \propto \mathrm{Cos}^2[\pi \underline{x}/(2a)], |\underline{x}|<a; \ 0, otherwise . \tag{20'}$$

This expression was chosen (for example, rather than a square pulse) to avoid the distracting artifacts from propagation of discontinuous edges. Also, numerical integration in *k*-space is somewhat simplified, avoiding complications where convergence depends on functions "oscillating to zero". From the Fourier transform of $\underline{\psi}[\underline{x}]$, the

temporally evolving configuration-space wavefunction $\psi[x,t]$ (as in Equation (14)) is given by:

$$\psi[x,t] \propto \int_{-\infty}^{+\infty} dk \frac{\text{Sin}[ka]}{k(1-k^2a^2\pi^{-2})} \omega_k^{-1/2} e^{i(kx-\omega_k t)}, \qquad (14')$$

which is easily computed numerically, as are first derivatives with respect to $x$ and $t$ for computing the densities $\rho_{NW}[x,t]$, $\rho[x,t]$, and current $J$.

In the following, the constant $a$ is taken to equal 1, so the initial *absolute* confinement is to within $|x| \leq 1$. Note, however, that with this specific initial state, confinement at the 95% level is within $|x| \leq 0.55$. This figure no doubt relates to the absolute initial Bohm-particle confinement given below.

Integrating $\rho_{NW}[x,t]$ results in the probability of obtaining an acausal NW position measurement, i.e., outside the light cone. Fig. 2 shows this probability vs time.

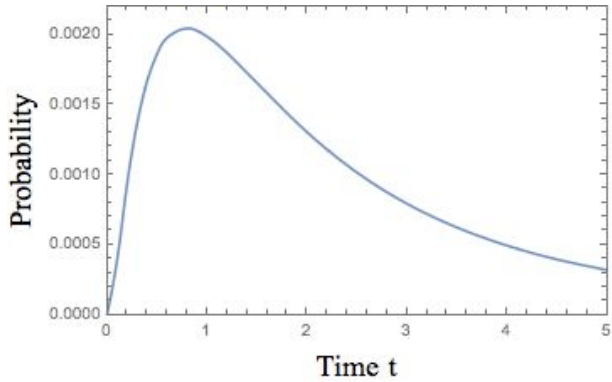

**Figure 1.** Probability of Newton-Wigner position measurement in space-like region vs time from initial localization within two Compton wavelengths.

With increased initial localization, this probability increases, while the probability density becomes more sharply located just ahead of fronts propagating at the speed of light.

It is now possible to compute the integral of the equations of motion $I[x,t]$ (Equation (12)) at any given values of $x$ and $t$. Integration simplifies since the real part in the principal-value integration in $I[x,t]$ vanishes by symmetry, leaving only the imaginary part with no singularity (and no principal-value complications in the numerical integration). An example is shown in Fig. 2 at $t = 0$.

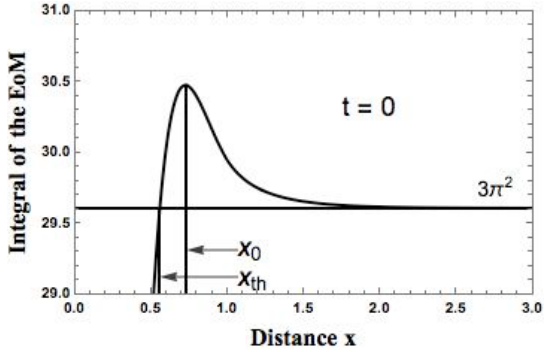

**Figure 2.** The integral of the equations of motion $I[x,t]$ plotted at $t = 0$. At $x < x_{th}$, a threshold value, position is a single-valued function of the integral, but becomes 2-valued above, corresponding to virtual-like pairs. The point $x_0$ increases with time and becomes a curve where pair annihilation occurs.

Contours of $x[t]$ can now be found where $I[x,t]$ is constant. The result is shown in Figure 3. In this case 815 values of $I[x,t]$ were computed over the range of $x$ and $t$ in a quarter of the figure (the double integration requiring about 3 minutes of laptop computing each). Contours were then simple to determine from the array of values.

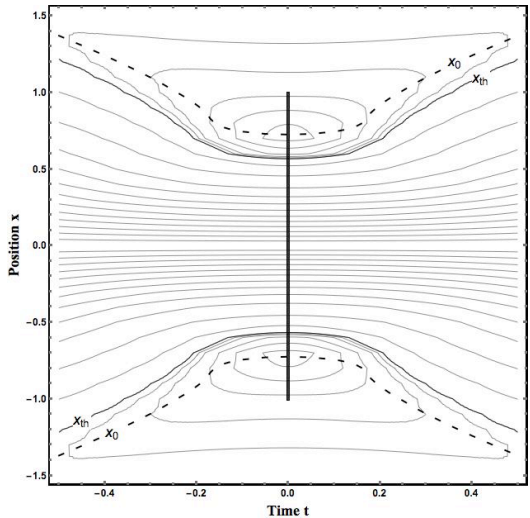

**Figure 3.** Bohmian virtual pair annihilation at an exploding front in the acausal spreading of a spin-0 wavefunction with Newton-Wigner position initially entirely localized to $|x| < 1$ (depicted as the solid vertical line at $t = 0$). Between the solid (threshold) and dashed annihilation curves, only virtual particles exist; beyond the dashed curves, only virtual anti-particles.

[The trajectories can also be understood more analytically, yet in approximation. At any time $t$ as in Fig. 3, near where the density $\rho$ changes sign, both $\rho$ and the current $J$ can be approximated as linearly depending on the distance from their separate zeros. Then $dx/dt = J/\rho$ can be solved for $x[t]$ in terms of the Lambert W (or "product log") function on two of its sheets, resulting in curves near pair annihilation in agreement

with Fig. 3.]

The system of Fig. 3 provides several details. First of all, advancing Bohmian real particles are found between $x = 0$ and a threshold value $x_{th}$ (= 0.57 at t = 0 for the $\cos^2$ system). Between $x_{th}$ and $x_0$ where $\rho$ changes sign ($x_0 = 0.72$ at t = 0), only virtual particles are present. These particles annihilate later with the purely virtual anti-particles exclusively found earlier at $x > x_0$ (in fact, at infinity for the virtual anti-particle paired with particle near $x_{th}$). Note that as only virtual pairs exist at $x > x_{th}$, the integral of $\rho$ from $x_{th}$ to $\infty$ (becoming negative at $x_0$) vanishes, whereas the integral of $\rho$ from 0 to $x_{th}$ equals the integral of $\rho_{NW}$ from 0 to $\infty$ (equal to 0.5). Incidentally, note that annihilation is not a Lorentz-invariant event. Lorentz transform moves the event to a separate space-time point.

It is as yet an open question whether a measurement $x_{NW}$ can be represented as a function of the sometimes 2-valued Bohmian trajectory, fixed time $t$, and their local fields in a way similar to Equation (36). If such a relation exists between $x_{NW}$ ($\vec{x}$ above) and the "primordial" Bohmian (sometimes virtual) particle trajectories as in Fig. 3, it must not be as simple as in the lowest relativistic case of the previous section.

Even a "0-th order" approximation is difficult. One possibility is that the NW measurement ignores the anti-particles. Superluminal expansion of the NW position in approximating the distribution of virtual particles in terms of simply the density $\rho$ then occurs since the extreme virtual particle positions expand in time as shown in Fig. 4.

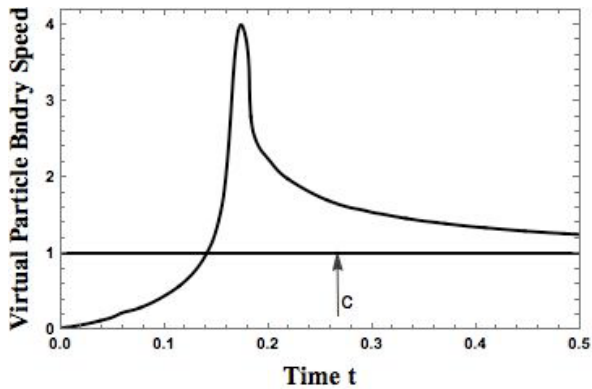

**Figure 4.** The virtual particle boundary (the dashed curves in Fig. 3) speed vs time t.

The virtual particle boundary speed increases in time from zero, reaches a maximum, and then asymptotically drops to the speed of light from above. This is qualitatively (though not quantitatively) consistent with Fig. 2. Unlike the exact NW values, the approximate positions are strictly finite for all $t > 0$.

Another possibility is that since $x_{NW}$ values are found for all $x > x_0$ at $t > 0$ perhaps the

$x_{NW}$ measurement may be "charge"-blind in sampling *both* Bohmian anti-particles and particles. This is an opposite extreme from ignoring the anti-particles and does lead to $x_{NW}$ values for all x at t > 0.

Speculation is perhaps simplest beginning with the situation at $t = 0$, since $v$ is everywhere initially zero. Note, however, that even though $v = 0$ initially, the system cannot be considered non-relativistic. Acceleration is so great that $|v|$ becomes large (in fact, infinite beginning from a single point where the invariant mass is zero) instantly as $t$ increases from 0. The 1-st order correction in Equation 29 is in the correct direction, but overcorrects.

Natural candidates then for a 0-th order approximation to the NW position in the case of the paired trajectories are the various relativistic centers of mass[22-24] in the asymptotic region $x > 1 + ct$ of interest considering the superluminal behavior[7] of the (orthodox) NW position. At $t = 0$, the centers of relativistic and invariant mass (amplitudes) are equal because of vanishing initial velocities. A further simplification with the state at $t = 0$ is the fact that the lab system is a center-of-momentum coordinate system, thus avoiding complications at $t > 0$ with Lorentz transforms back and forth to a center-of-momentum system for defining the centers of mass.

Though the exact NW position is bounded at $t = 0$, the center of mass amplitude is unbounded because of the trajectory at infinity. The relativistic mass amplitude can be shown to fall off with anti-particle position $x_A$ only as $x_A^{-\frac{1}{2}}$, so that used as a weighting on $x_A$, contribution to a center of mass amplitude *increases* as $x_A^{+\frac{1}{2}}$. This may not be problematic as an approximation, since the integral of motion can be shown to approach a constant (equal to $3\pi^2$, in fact, for all time t) asymptotically with distance exponentially as seen in Fig. 2. This results in the antiparticle position $x_A$ dropping from infinity rapidly (logarithmically) to the annihilation value $x_0$ as the paired particle position $x$ increases from the virtual-particle threshold $x_{th}$, and the probability distribution for the approximate $x_{NW}$ approaches zero rapidly with increased $x_{NW}$ (as a Gaussian function). However, a gap therefore exists between $x_{th}$ and $x_0$. A 0-th order approximation then requires a constant shift from the center of mass amplitude. The result is shown in Fig. 5.

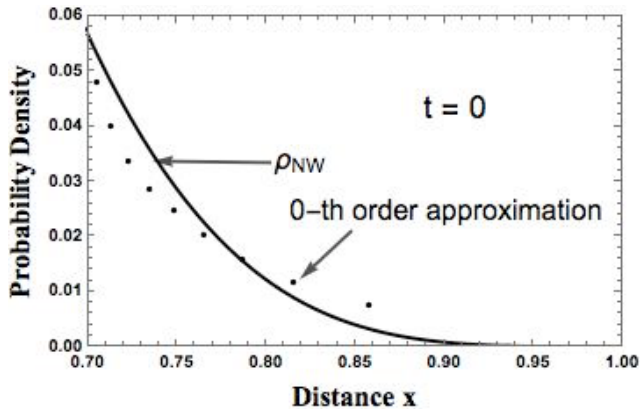

**Figure 5.** Probability density for observing the NW particle position (solid curve) approximated (solid dots) as the center of paired virtual-particle mass for an ensemble of particles in their initial positions.

Another possibility for a 0-th order approximation is the center of mass[2], which is finite for all trajectories and in fact leads to improvement over the result shown in Figure 5 with the approximate $x_{NW}$ strictly limited at t = 0. Preliminary work on extending the above to $t > 0$ indicates asymptotic decrease in the probability of apparent superluminal motion in center-of mass-approximations similar to Fig. 2 with the exact NW position. Details are yet to be fleshed out.

**Discussion**

As seen in Fig. 3, this system is quite peculiar considering the rarified closed loops away from the origin. Evidently, the initial state is so contorted that some position measurements correspond to particles that would disappear if left alone. However, considering the non-local aspect of the Newton-Wigner position $\underline{x}_i$ as indicated by the above near-nonrelativistic calculation, it is conceivable that $\underline{x}_i$ at time *t,* in depending on the entire trajectory, could be non-vanishing even with annihilation occurring before *t*. There seems no hint in the mathematics that the loops are connected in some way to the more ordinary trajectories shown. Therefore we have treated the loops as independent paired particles comprising a small part of a larger ensemble.

Then an interesting Bohmian interpretation of the peculiarities of the Newton-Wigner position is suggested. The "something" not seen outside the localized pulse at *t* = 0 is a small set of virtual Bohmian anti-particles. Furthermore, spatial extension beyond isolated point particles requiring space-like hyperplane specification for covariance as suggested in Ref. 7 relates to the spatial separation between virtual particle and paired antiparticle. This separation may not be at odds with current experimental evidence for point-particle character of an electron for example, where pairs may be absent.

Not only is the observation from a Lorentz-boosted frame "explained", but also, acausality relates to both superluminality near Bohmian pair annihilation and pair separation. As temporal evolution proceeds, the virtual (i.e., "hyper-entangled") particles and anti-particles annihilate each other at the propagating pulse-edge fronts. The probability of obtaining a Newton-Wigner position value in the space-like regions then decreases with time (Fig. 1) as the virtual anti-particles are swept up.

It might be thought that the multi-particle pair existence and annihilation seen in Fig. 3 would be questionable without interaction with an external field as quantified in quantum field theory. However, the theory considered here is strictly confined to single-particle quantum mechanics and allows *no* pair creation or annihilation at the orthodox level. At any time t, the NW position is always single-valued and is described statistically via $\rho_{NW}$, a conserved positive-definite function.

Orthodox quantum mechanics makes *no* predictions at the Bohmian sub-quantum or hidden-variable level. However, even at this level the theory considered here covers no *free* anti-particles. On the contrary, any Bohmian pairs are comprised of *virtual-like* particles that recombine within times of the order of the (Compton wavelength) / c. At this sub-quantum level, there is only a hint of a multi-particle theory that a generalization to quantum field theory could provide.

Extending the ideas of this paper to particles with spin may also be possible. The choice taken by most researchers for a Bohmian theory of single spin-1/2 particles is to assume the velocity is given in terms of the conserved current $\bar{\psi} \gamma_\mu \psi$; see, for example, Ref. 15. This approach would include particle trajectories covering internal aspects. For example, with a superposition of two positive-energy spin-1/2 plane waves of different frequencies with spins aligned and normal to momentum, particles oscillate transversely to both spin and momentum with |v| repeatedly often reaching the speed of light c momentarily while advancing at small constant velocity along the momentum.

Alternatively, particle velocity may be taken from the conserved convective part of the spin-1/2 current in its Gordon decomposition (regardless of Dirac's original intent of positive probability densities for solely particles). The internal current associated with zitterbewegung and with the curl of the magnetic moment density is then separated off. This convective-current theory is then on the same footing as the above spin-0 theory. Steps in this direction are presented in Appendix B.

Aside from further research into relating dBB to orthodox quantum theory in the relativistic limit, other directions are also suggested. Peculiarities of the dBB theory's treatment of massless fields[27], possibly connected to bound-state particle stasis, must be unraveled. Though major effort has already been devoted to a dBB underpinning of quantum field theory,[12-15,25,26] perhaps the peculiar localized virtual-like multiplicity of particles seen here (e.g., in Fig. 3) corresponding to what is a single particle in the non-relativistic regime may be suggestive of a detailed picture of the creation and annihilation of real particles. The hidden multiplicity associated with a single-particle quantum mechanics, however, is very limited—all particles are completely separated spatially by surfaces of $\rho = 0$. An interacting-field theory may provide connection between real particles and, of course, an understanding of the difficult measurement problem. A deeper theory of entanglement may result from what appears as acausal in the free-field case.

More ambitious is an understanding of the range of possible deterministic theories which reproduce quantum mechanics. The dBB scheme can hardly be the only possibility. At present the theory can be regarded mainly as an example as to how deterministic particles could exist in harmony with quantum-mechanical fact.

On the other hand, in dodging Occam's razor occasionally or at least until the dBB idea is understood more fully than at present, the theory may provide an understanding. It may be useful to conceive of quantum mechanical phenomena *as if* particles were moving in such and such a manner, as in explaining peculiarities of the Newton-Wigner particle localization presented above.

**Acknowledgement** The author thanks Gordon Fleming for many discussions about Newton-Wigner localization.

# Appendix A. Discrete plane-wave superposition*

The integral *I*, expressed as sums for the superposition of a finite number of discrete plane waves, is particularly useful for computing particle trajectories, which may be determined by calculating contours from the resulting transcendental equation. Equation (13) for the discrete case reads:

$$\psi[z,t] = \sum_{k'} \omega_{k'}^{-1/2} \phi_{k'} e^{ik'z - i\omega_{k'}t} . \tag{13'}$$

Equation (10) then becomes, in terms of the expected value $\langle k/\omega_k \rangle c^2$ of the group velocity:

$$v/c^2 = \frac{\langle k/\omega_k \rangle + \tfrac{1}{2} \sum_{k''\neq k'} \sum (\omega_{k'}\omega_{k''})^{-1/2} \phi_{k'}^* \phi_{k''} (k''+k') \, Exp[i(k''-k')z - i(\omega_{k''} - \omega_{k'})t]}{1 + \tfrac{1}{2} \sum_{k''\neq k'} \sum (\omega_{k'}\omega_{k''})^{-1/2} \phi_{k'}^* \phi_{k''} (\omega_{k'} + \omega_{k''}) \, Exp[i(k''-k')z - i(\omega_{k''} - \omega_{k'})t]} . \tag{10'}$$

Finally, Equation (12) translates to:

$$I = i(z - \langle k/\omega_k \rangle c^2 t) + \tfrac{1}{2} \sum_{k''\neq k'} \sum (\omega_{k'}\omega_{k''})^{-1/2} \phi_{k'}^* \phi_{k''} (\omega_{k'} + \omega_{k''}) (k''-k')^{-1} \, Exp[i(k''-k')z - i(\omega_{k''} - \omega_{k'})t] . \tag{12'}$$

Trajectories $z[t]$ are then easily computed numerically from Equation (12') at $I$ = constant, noting that the double sum in Equation (12') is pure imaginary.

Figure A-1 illustrates trajectories for an ensemble of particles at rest in the mean resulting from a superposition of three plane waves, two of which are ultra-relativistic. Continual creation and annihilation of pairs in the concept of Feynman[8,9] and Stückelberg[10] of anti-particles as particles moving backwards in time is evident. The particles are *virtual* in the sense that the pairs are always maximally entangled; any pair created is quickly annihilated. Creation or annihilation of free particles is not possible here where no interaction with other fields is permitted. Note that there exist situations where the anti-particles are not simply apparent, i.e., cannot be removed by proper Lorentz transformation. Diagonals of the figure form the light cone, and the position-axis length is about 0.01 Compton wavelengths.

---

* The previously unpublished research in Appendices A and B was completed in 1988 and 2004, respectively.

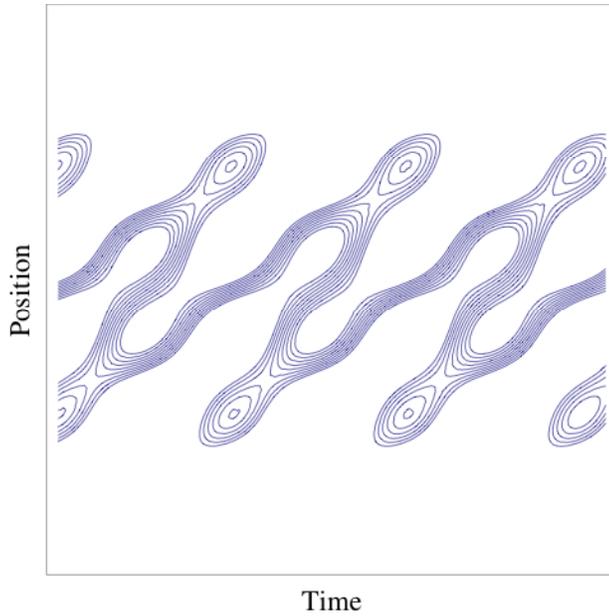

**Figure A-1.** Particle trajectories in the mean rest frame corresponding to the superposition of three extraordinary plane waves, two of which are ultra-relativistic. Diagonals of the figure specify the light cone and the vertical axis is of the order of 0.01 Compton wavelength. Note the peculiar virtual multi-particle aspect as a result of an extreme pilot wave's effect on a single-particle existence.

## Appendix B. Free Spin-1/2 Particles

**generalities**

The Bohmian spin-1/2 theory is considerably more complicated than for spin-0. With the extra degrees of freedom in the spinor wavefunction, it is not obvious what the representation analogous to Equation (2) would be, and therefore what sort of Lagrangian or Hamilton-Jacobi equation would result. Nevertheless, by adopting a small part of the scalar field results, a complete free spin-1/2 single-particle theory is possible.

The obvious choice for a Bohmian theory of single spin-1/2 particles is to assume the velocity is given in terms of the conserved current $\bar{\psi}\gamma_\mu\psi$.[13] This approach would include particle trajectories covering internal aspects. For example, with a superposition of two positive-energy spin-1/2 plane waves of different frequencies with spins aligned and normal to momentum, particles oscillate transversely to both spin and momentum with |v| repeatedly often reaching the speed of light c momentarily while advancing at small constant velocity along the momentum. Many other states' particles oscillate close to the speed of light. Such states have been described[28] as comprising a set of measure 0 relative to the set of possible states; of course, a measure-0 set depends on the measure, and furthermore some such sets can be significant or interesting.

Here, however, we focus on the conserved convective part of the spin-1/2 current in its Gordon decomposition (regardless of Dirac's original intent of positive probability densities for solely particles). The internal current associated with zitterbewegung and with the curl of the magnetic moment density is then separated off. This approach with convective current then parallels the above spin-0 theory.

Further, allow for the possibility of an effective rest mass function $\mu_0[x]$. Then Equations (3) and (5) with the scalar theory suggest (sidestepping Equation (2)):

$$\mu_0 u_\mu = \frac{\hbar}{2i\bar{\psi}\psi}\left[\bar{\psi}\frac{\partial\psi}{\partial x_\mu} - \frac{\partial\bar{\psi}}{\partial x_\mu}\psi\right], \tag{B-1}$$

but where now the spinor $\bar{\psi} = \psi^\dagger\gamma_0$, $\psi$ solves the free Dirac equation, and $\gamma_0$ is

$$\gamma_0 = \begin{bmatrix} +1 & 0 & 0 & 0 \\ 0 & +1 & 0 & 0 \\ 0 & 0 & -1 & 0 \\ 0 & 0 & 0 & -1 \end{bmatrix}. \tag{B-2}$$

Since $u_\mu u^\mu = -c^2$, Equation (B-1) immediately expresses the effective invariant mass $\mu_0$ in terms of $\psi$ by:

$$(\mu_0 c)^2 = -\mu_0 u_\mu \cdot \mu_0 u^\mu. \tag{B-3}$$

However, an exact analogue of Equation (8) is obtained by defining a specific spin tensor in addition to the quantum potential in terms of the Dirac spinors. As with the scalar field, let

$$\Phi \equiv -\frac{\hbar^2}{2m_0}(\bar{\psi}\psi)^{-1/2}\ \Box(\bar{\psi}\psi)^{1/2}. \tag{B-4}$$

Furthermore, define an improper (in the sense of wavefunction in the denominator) tensor $T_{\mu\nu}^{spin}$ as:

$$T_{\mu\nu}^{spin}\hbar^{-2} \equiv \tfrac{1}{2}(\bar{\psi}\psi)^{-1}\left[\frac{\partial\bar{\psi}}{\partial x_\mu}\frac{\partial\psi}{\partial x_\nu}+\frac{\partial\bar{\psi}}{\partial x_\nu}\frac{\partial\psi}{\partial x_\mu}\right]$$
$$-\tfrac{1}{2}(\bar{\psi}\psi)^{-2}\left[\left(\frac{\partial\bar{\psi}}{\partial x_\mu}\psi\right)\left(\bar{\psi}\frac{\partial\psi}{\partial x_\nu}\right)+\left(\frac{\partial\bar{\psi}}{\partial x_\nu}\psi\right)\left(\bar{\psi}\frac{\partial\psi}{\partial x_\mu}\right)\right]. \tag{B-5}$$

The symmetric second-rank tensor field $T_{\mu\nu}^{spin}$ has the interesting property that it is invariant under the replacement, $\psi \to f[x]\cdot\psi$, where $f[x]$ is an arbitrary complex-valued function of space-time $x$. In particular, $T_{\mu\nu}^{spin}$ is independent of an overall amplitude or phase, and would vanish if $\psi$ were a scalar.

In terms of $\Phi$ and $T_{\mu\nu}^{spin}$, Equation (B-3) reads:

$$(\mu_0 c)^2 = (m_0 c)^2 + 2m_0\Phi + tr[T^{spin}], \tag{B-6}$$

similar to Equation (8) aside from the final term. Equation (B-6) follows from the fact that the four components of $\psi$ each solve the GKS equation. The proof is entirely straightforward and can be checked using a program capable of symbol manipulation.

The equations of motion can now be obtained from

$$\mu_0 d(\mu_0 u_\mu)/d\tau = \mu_0 u^\nu\ \partial(\mu_0 u_\mu)/\partial x_\nu \tag{B-7}$$

by expressing the right-hand side in terms of $\psi$ by using Equation (B-1). Remarkably, the resulting equations of motion can be expressed in terms of $\Phi$ and $T_{\mu\nu}^{spin}$ as simply:

$$\mu_0 u^\nu\ \partial(\mu_0 u_\mu)/\partial x_\nu = -\frac{\partial m_0\Phi}{\partial x_\mu}-\frac{1}{\bar{\psi}\psi}\partial^\nu\left[\bar{\psi}\psi\ T_{\mu\nu}^{spin}\right], \tag{B-8}$$

similar to Equation (9), except for stress tensor $\bar{\psi}\psi\ T_{\mu\nu}^{spin}$. The proof of Equation (B-8), though straightforward, has an excessive number of steps, but can be checked using a program capable of symbol manipulation.

**non-relativistic limit**

We now specialize to positive-energy solutions of the Dirac equation by using the Foldy-Wouthuysen representation with 3$^{rd}$ and 4$^{th}$ components of the spinor set equal to zero. An arbitrary such spinor with four real components can be represented in terms of an amplitude $A$ and phase $S$ as

$$\psi = A e^{\frac{iS}{\hbar}} u(\hat{s}), \tag{B-9}$$

where $\hat{s}$ is a 3-dimensional $x$-dependent unit spin pseudovector field, and where $u(\hat{s})$ with $u^\dagger u = 1$ and $s_k = u^\dagger \sigma_k u$ is:

$$u(\hat{s}) = \frac{1}{\sqrt{2(1+s_3)}} \begin{bmatrix} 1+s_3 \\ s_1 + is_2 \\ 0 \\ 0 \end{bmatrix}. \tag{B-10}$$

Equation (B-9) is of course the spin-1/2 analogue of the Bohm-de Broglie expression, Equation (2), with $u(\hat{s})$ to account for the extra two degrees of freedom. In terms of the amplitude $A$, the quantum potential is again given by Equation (7). Furthermore, a straightforward calculation indicates that the tensor $T^{spin}_{\mu\upsilon}$ is given simply by:

$$T^{spin}_{\mu\upsilon} = \frac{\hbar^2}{4} \frac{\partial s_l}{\partial x_\mu} \frac{\partial s_l}{\partial x_\upsilon}. \tag{B-11}$$

The non-relativistic limits can now be given. The quantum potential $\Phi$ is given by

$$\Phi = -\frac{\hbar^2}{2m_0} A^{-1} \Delta A, \tag{B-12}$$

where $\Delta$ is the Laplacian operator, and the spin tensor $T^{spin}_{ij}$ is approximately:

$$T^{spin}_{ij} = \frac{\hbar^2}{4} \frac{\partial s_l}{\partial x_i} \frac{\partial s_l}{\partial x_j}. \tag{B-13}$$

Interestingly, this tensor is identical to an earlier result found[11] for a non-relativistic spinning object, although the particle in the theory presented here is not taken as spinning; rather, the pilot wave has a spin component that affects where the particle moves, for example in a Stern-Gerlach experiment.

The ensemble equations of motion become:

$$\frac{\partial v_j}{\partial t} + \frac{\partial v_j}{\partial x_i} v_i = -m_0^{-2}\left(\frac{\partial m_0 \Phi}{\partial x_j} + \frac{1}{A^2}\frac{\partial}{\partial x_i}\left[A^2 T_{ji}^{spin}\right]\right), \tag{B-14}$$

similar to the Navier-Stokes equation for a viscous fluid, except that here dissipation and anti-dissipation cancel in the mean.

Note that since

$$\int d^3x \ A^2 \frac{\partial}{\partial x_j} \frac{\Delta A}{A} = -2\int d^3x \frac{\partial A}{\partial x_j} \Delta A$$
$$= 0 \ , \tag{B-15}$$

Equations (B-12) and (B-14) give the ensemble average $<dv_j/dt>$ at any fixed time as:

$$\int d^3x \ A^2 \frac{dv_j}{dt} = 0. \tag{B-16}$$

Furthermore, even in the non-relativistic limit, the ensemble fluid is rotational, with circulation given by:

$$curl_k v = \frac{\hbar}{4m_0} \varepsilon_{kji}\varepsilon_{lmn} s_l \frac{\partial s_m}{\partial x_j}\frac{\partial s_n}{\partial x_i} . \tag{B-17}$$

Particles with spin therefore exhibit richer motion than scalar particles. The tensor component of the force of wave on particle results in a circulatory ensemble fluid.